\documentclass{article}
\usepackage[utf8]{inputenc}
\usepackage{geometry}
\usepackage{graphicx}
\usepackage{bm}
\usepackage{amsmath}
\usepackage[normalem]{ulem}
\usepackage{graphicx}
\usepackage{color,soul}
\usepackage[colorlinks=true, citecolor=red, urlcolor=blue]{hyperref}
\usepackage{tabularx}
\usepackage{textcomp}
\usepackage{graphicx}
\usepackage{adjustbox}
\usepackage{authblk}
\usepackage{caption}
\usepackage[table,xcdraw]{xcolor}
\usepackage{caption}
\usepackage{booktabs}
\usepackage[backend=biber, style=chem-acs, articletitle=true]{biblatex}
\usepackage{float}
\usepackage{longtable}
\renewcommand*{\cite}{\autocite}

\begin{document}

\title{Field-Driven Hybrid Filament Formation Governs Switching in Ta–HfO$_2$–Pt Memristors}

\author[1,2]{Ashutosh Krishna Amaram}
\author[3]{Aditya Koneru\textsuperscript{*}}
\author[1,2]{Subramanian KRS Sankaranarayanan\textsuperscript{**}}

\affil[1]{Department of Mechanical and Industrial Engineering, University of Illinois, Chicago, Illinois 60607, United States}
\affil[2]{Center for Nanoscale Materials, Argonne National Laboratory, Lemont, Illinois 60439, United States}
\affil[3]{Argonne Leadership Computing Facility, Argonne National Laboratory, Lemont, Illinois 60439, United States}
\affil[ ]{\textsuperscript{*} E-mail: akoneru@anl.gov}
\affil[ ]{\textsuperscript{**} E-mail: skrssank@uic.edu}
\maketitle
\begin{abstract}
Memristive devices have gained significant attention for their potential in next-generation non-volatile memory and neuromorphic computing architectures. Among emerging candidates, transition metal oxides have proven particularly promising. While the switching mechanism in Ta/HfO$_2$/Pt devices was long attributed solely to oxygen vacancy based filaments, recent experimental evidence suggests a more complex dual-regime: the diffusion of metal cations also contributes to the formation of a conductive bridge. However, the precise atomistic mechanisms governing this metal cation migration remain poorly understood. Additionally, the role of defects such as oxygen vacancies present in the transition metal oxide in determining the final filament size and shape is also not well understood. Here, we employ molecular dynamics (MD) simulations with dynamic charge transfer to provide a detailed analysis of the atomistic mechanisms governing the co-formation of Ta-cation and oxygen-deficient filaments. We clearly show how varying the initial oxygen vacancy concentrations and spatial configurations within the HfO$_2$ matrix influences the final morphology and dimensions of the conductive filament. The switching is governed by field-driven formation and rupture of a hybrid Ta-cation-rich, oxygen-deficient filament in HfO$_2$. Our simulations closely match experiment, validating the model as a robust framework for understanding switching in oxide memristors and guiding designs that reduce cycle-to-cycle and device-to-device variability—key barriers to high-performance devices.

\noindent 
\end{abstract}

\newpage

\section{Introduction} 
Memristors have gained a lot of attention owing to their applications in non-volatile memories~\cite{non-volatile-memory1,non-volatile-memory2,non-volatile-memory3} and human-brain inspired neuromorphic computing architectures~\cite{park2022dynamic_memristor_neuron, aguirre2024hardware_hardwar_implement}. The device architecture of a memristor is a simple metal-insulator-metal (MIM) stack where a variety of semiconducting materials, both transition metal oxides~\cite{oxide_memristors} and 2-dimensional materials~\cite{2dmemristors}, have been tested and have shown promising results in the form of low voltage consumption~\cite{Ge2018Atomristor}, fast-switching speeds~\cite{Nibhanupudi2024UltraFast, Bottger2020Picosecond}, and high endurance and retention characteristics~\cite{Yuan2025HighEndurance}. These characteristics make memristors highly attractive for realizing high-density integration of electronics~\cite{Jayachandran2024_3DIntegration}. The physical mechanism associated with switching of different memristive devices is a characteristic of the semiconducting material. Transition metal oxides like VO$_2$ switch because of an insulator-metal transition that is triggered when the device reaches the set voltage and the phase transition temperature~\cite{Schmid2024Picosecond}. In complex perovskites like PCMO, switching is mainly caused by charge trapping at the metal-semiconductor interface~\cite{PCMO}. However, the majority of semiconductors which have been explored for memristive applications form a filament that acts like a conductive bridge between the two electrodes. Depending upon the composition of the filament, this class of filamentary based memristors is divided into two sub-categories. One is valence-charge-memory (VCM)~\cite{Kaniselvan2023Atomistic,VCM-1,VCM-2,VCM-3,VCM-4,VCM-5} where the metal electrodes are mostly inert in nature like Pt and the insulator materials are transition metal oxides like HfO$_2$~\cite{Zhang2021HfO2Filament}, TiO$_2$~\cite{tio2} and Ta$_2$O$_5$~\cite{Kang2015MultilevelTaOx}. In this class of memristors, the primary switching mechanism for High Resistance State (HRS) to Low Resistance State (LRS) switching is the formation of a positively charged oxygen vacancy filament between the two electrodes. The other category of filamentary based memristive devices is electrochemical metallization memories (ECM)~\cite{Menzel2013Switching,ECM-1,ECM-2,ECM-3}. In this class, the memristor stack is composed of highly diffusible metal electrodes like Cu, Ag, Al, Ni etc. and the insulator is composed of transition metal oxides like HfO$_2$, TiO$_2$ and Ta$_2$O$_5$ or 2-dimensional materials like MoS$_2$~\cite{Ran2025FilamentDynamics,MoS2-1,MoS2-2,MoS2-3,MoS2-4,MoS2-5,MoS2-6} or hBN~\cite{Volkel2023hBNMemristors,hBN-1,hBN-2,hBN-3,hBN-4}. In these devices, the primary switching mechanism is governed by the formation of a metallic filament between the two electrodes.
\vspace{0.5em}

Recent advancements in memristive devices have revealed a complex switching behavior governed by the simultaneous migration of metal cations and oxygen-deficient regions. Experimental studies using Scanning Tunneling Microscopy (STM) have confirmed that metal cation motion complements oxygen anion movement in AO$_x$ devices (where A = Ta, Ti, or Hf)~\cite{Wedig2016NanoscaleCation}. Specifically, investigations into Ta/HfO$_2$/Pt hetero-structures using Scanning Transmission Electron Microscopy (STEM) and Electron Energy Loss Spectroscopy (EELS) have directly visualized conductive bridges characterized by both Ta-cation enrichment and oxygen vacancy concentration~\cite{Jiang2016Sub10nmTa}. These findings reveal that earlier models with a simplified theory of only oxygen vacancies participating in the switching mechanism are not sufficient and a hybrid model is required to understand the switching characteristics of the memristive device which would integrate the migration of both oxygen vacancies and metal cations. Despite previous experimental observations, a step-by-step physical mechanism governing the formation and dissolution of these hybrid filaments remains elusive. In particular, the temporal sequence of the participating species—that is, whether the oxygen vacancies are first nucleated to create a favorable pathway for metal cation diffusion or these cations diffuse first across the semiconducting layer causing the oxygen anions to migrate towards the electrode region leaving behind oxygen vacancies—remains unclear. Furthermore, there is a critical lack of research correlating the initial stochastic distribution of oxygen vacancies with the resulting filament geometry. These vacancies might arise while depositing the material or during the fabrication process. The distribution of these defects is not deterministic in nature and is rather randomly arranged; thus, the role of such defects in governing the geometric and electrical parameters of the conductive filament must be understood to fine-tune the processing parameters of device fabrication and help the user gain a better control over device performance. Addressing this gap is essential, as cycle-to-cycle and device-to-device variability remain the primary bottlenecks preventing the large-scale commercialization of memristive technology, particularly in the use of these memristive devices in non-volatile memories and neuromorphic computers. To overcome this challenge, a robust modeling framework is required where one can simulate different device configurations and study the entire kinetic process governing the switching mechanism of the device. Moreover, the modeling framework should be able to quantify the main cause of variability, thus allowing one to fabricate more reliable memristive systems. By bridging the gap between atomistic insights and continuum scale parameters, we will have a powerful tool capable of predicting the performance of such devices with different configurations and operating conditions.
\vspace{0.5em}

Here, we present a comprehensive molecular dynamics based framework with dynamic charge transfer that reveals the governing mechanism of Ta rich filament formations in Ta-HfO$_2$-Pt memristive devices, which is further validated with charge maps and IV characteristics. Unlike previous approaches that had a pure continuum based modeling approach, our framework integrates atomistic scale trajectories produced by running MD simulations on devices exposed to different initial defects and electrical stimuli with IV characteristics. The IV characteristics are obtained by an analytical model that accounts for the number of Ta cations participating in the filament formation and rupture process. Thus, the model helps establish a coherent structure-property correlation for the device where filament geometry is carefully correlated with memristive properties like switching ratio, voltages, and the nature of the switching mechanisms (abrupt or gradual). We also explore the role of initial oxygen vacancy concentration and spatial configurations of such vacancies in HfO$_2$ on the switching characteristics of the device. Beyond binary resistive switching mechanism, we also put our device to the test by applying a pulsed field cycle in order to understand the potentiation and depression characteristics of the device, which is critical in order to test the capability of the device to reproduce the necessary conductance states that would be used as synaptic weights in performing computing tasks with a neuromorphic computer. This test was also performed across devices with varying vacancy concentration to assess the stochasticity associated with the device hindering its commercialization in exciting applications. Overall, our simulations not only elucidate the atomistic mechanism governing the kinetics of filament formation and the dissolution process but also extend their capability in assessing device functionality along with revealing the cause of variability in such devices.  

\section{Computational Details and Methodology}

\subsection{Device Set-up and Atomistic Simulation Framework}

\begin{figure}[htbp]
    \centering
    \includegraphics[width=1\textwidth]{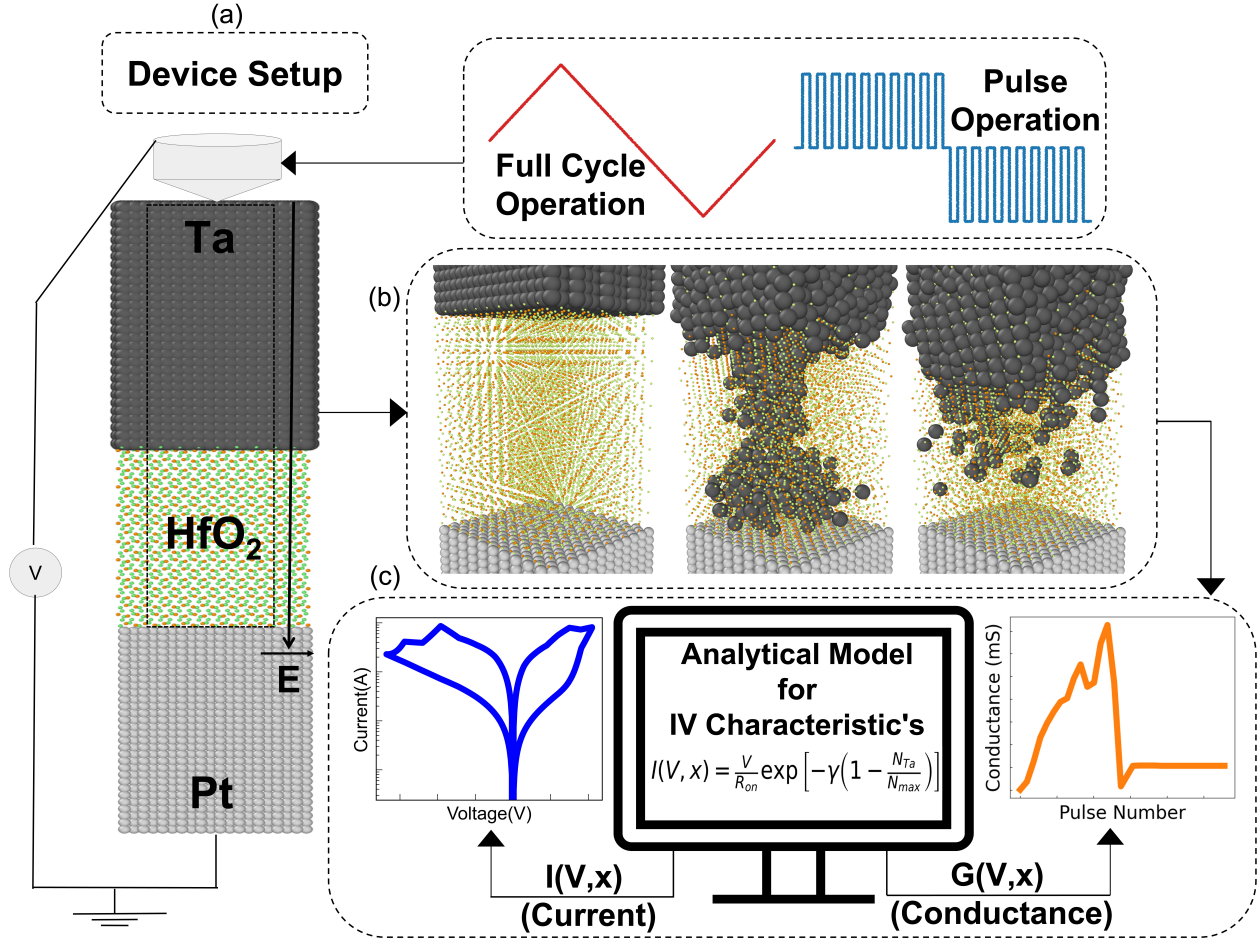}
    \caption{\textbf{An end to end atomistic simulation framework to understand switching mechanism in Ta-HfO$_2$-Pt memristive devices}.(a) Shows the MD setup of the device where pristine and defective(0.81\% and 1.35\% oxygen vacancies) were studied under two different electrical stimuli:(i) full cycle field operation and (ii) field pulse operation. (b) shows the MD trajectories of the device when studied under full cycle field operation. (c) The MD trajectories are then used to quantify the number of Ta cations participating in the filament formation and dissolution process. This information is then fed as an input to an analytical model to obtain an IV characteristic for a device under full cycle field operation. For the same device under field pulse operation the model helps us to first compute conductance which then is used to obtain a conductance versus pulse number plot highlighting the neuromorphic behavior of the memristive device.}
    \label{fig:method}
\end{figure}

We develop a detailed atomistic simulation framework which first performs molecular dynamics (MD) simulations and then uses the MD trajectories to further compute the atomic charge distribution, stoichiometry, IV characteristics, and filament conductivity as shown in Figure~\ref{fig:method}. A detailed understanding of the analytical modeling of IV characteristics is explained later in a separate subsection.
\vspace{0.5em}

The MD simulations were performed using the LAMMPS molecular dynamics simulator~\cite{Thompson2022LAMMPS}. Periodic boundary conditions were applied along the x and y directions, while non-periodic boundary conditions were imposed along the z direction. A time step of 3 fs was used for all MD simulations. An electric-field cycle was applied from 0 to 1.4 to 0 to -1.4 to 0 V/Å with a ramp rate of 0.022 V/Å as shown in Fig. 2. The system temperature was maintained at 300 K using the NVT ensemble during the set cycle. During the reset cycle, the temperature of the filament region was increased to 900 K using the Berendsen thermostat to mimic Joule heating and accelerate the filament rupture process, while the remainder of the system was maintained at 300 K. For pulse operation, the amplitude for the positive pulse was fixed at 1.4 V/Å and for the negative pulse the amplitude was fixed at -1.4 V/Å. The pulse width was fixed at 30 ps as shown in Fig. 3. The temperature during the entire duration of pulse operation was maintained at 300 K using the NVT ensemble.

\subsection{Functional Form}
The interatomic interactions were modeled using the charge-transfer interatomic potential (CTIP)~\cite{Sasikumar2019MachineLearning, ctip-high-entropy, Lahkar2025TaOxHfO2ReRAM, Sasikumar2017CTIP}. This potential combines the embedded atom method (EAM) with an electrostatic term, capturing reactive simulations that are essential for understanding the atomistic behavior of electronic devices under an applied external electric field. Within the CTIP framework, metal–oxygen interactions are described by the electrostatic term, while metal–metal interactions are captured by the embedded atom method. The functional form of CTIP is shown in Eq.~\ref{eq:CTIP} where E$_{es}$ is the electrostatic term and E$_m$ is the embedded atom method term.

\begin{equation}
\label{eq:CTIP}
E_t = E_{es} + E_m
\end{equation}

The functional form of E$_{es}$ is represented by Eq.~\ref{eq:electrostatic_term} where the model imposes an upper and lower bound on the charges of individual cations and anions in order to avoid them from attaining charges that exceed their valence charges. q$_{min}$ and q$_{max}$ are the upper and lower bounds of the charge term q$_i$. There also exists a coefficient $\omega$ that would penalize the metal atom if it gains electrons or loses inner shell electrons. The coefficient also penalizes oxygen anions if they lose electrons or receive more than two electrons, which would otherwise exceed their limit of the number of valence electrons. X$_i$ represents the self energy of atom i and V$_{ij}$ represents the Coulombic interaction term between atom i and j.

\begin{equation}
\begin{aligned}
\label{eq:electrostatic_term}
E_{es} =\; & E_0 
+ \sum_{i=1}^{N} q_i \chi_i
+ \frac{1}{2} \sum_{i=1}^{N} \sum_{j=1}^{N} q_i q_j V_{ij} \\
& + \sum_{i=1}^{N} \omega
\left(
1 - \frac{q_i - q_{\min,i}}{\left| q_i - q_{\min,i} \right|}
\right)
\left( q_i - q_{\min,i} \right)^2 \\
& + \sum_{i=1}^{N} \omega
\left(
1 - \frac{q_{\max,i} - q_i}{\left| q_i - q_{\max,i} \right|}
\right)
\left( q_i - q_{\max,i} \right)^2
\end{aligned}
\end{equation}

The functional form of E$_m$ is represented by Eq.~\ref{eq:EAM_term}. This term is used to model the non-electrostatic interactions in the system. The term $\phi$$_{ij}$(r$_{ij}$) represents the pair-wise interactions between atom i and j separated by the distance of r$_{ij}$.

\begin{equation}
\label{eq:EAM_term}
E_m = \frac{1}{2} \sum_{i=1}^{N} \sum_{j=1}^{i_M} \phi_{ij}(r_{ij})
+ \sum_{i=1}^{N} F_i(\rho_i)
\end{equation}

In Eq-3 the term F$_i$ represents the embedding energy which is the energy required to embed an atom i onto a site with electron density $\rho$$_i$. A detailed expression of CTIP potential is described in the supplementary information.

\subsection{Analytical Modeling of IV Characteristics}
The IV characteristics of the device are obtained by constructing a direct physics-based correlation between the atomistic evolution of the Ta cation based filament and the macroscopic electrical response. MD simulations reveal that the governing mechanism for switching the device between LRS and HRS is the formation and dissolution of a Ta cation based filament. Therefore, we introduce a dimensionless state variable $x(V)$ defined in Eq.~\ref{eq:state_variable}.

\begin{equation}
\label{eq:state_variable}
x(V) = \frac{N_{\mathrm{Ta}}(V)}{N_{\mathrm{Ta,max}}}, \qquad 0 < x < 1
\end{equation}

Where $Na_{Ta}(V)$ denotes the number of Ta cations participating in the filament formation process when an external electric field is applied across the device. $Na_{Ta,max}(V)$ denotes the maximum population of Ta cations constituting the filament after the completion of the set process. By taking the ratio of these two quantities, we ensure that the variable $x$ is physically constrained between 0 and 1 where 0 represents the ruptured state of the filament and 1 represents the formed state of the filament. In order to map the atomistic information we obtain from MD simulations with the macroscopic electrical response, we assume that the resistance of the device increases exponentially with filament constriction and rupture. The functional form of the device resistance is described by Eq.~\ref{eq:resistance}. 

\begin{equation}
\label{eq:resistance}
R(x) = R_{\mathrm{on}} \, e^{\gamma (1 - x)}
\end{equation}

This functional form of resistance is physically justified by the atomistic response of the device to an external electric field. During the set process, a filament composed of Ta cations is nucleated and it grows in a continuous fashion until the completion of the set process. This process is captured by the state variable $x$ which would increase until it reaches a value of 1. At this state, $R(x)$ attains a value of $R_{on}$ which is defined as the low resistance state of the device. This equation also captures the entire process of reset with utmost accuracy. As the filament begins to constrict in size, the resistance exponentially increases due to enhanced atomic scale gaps that increase the tunneling barriers for electron conduction. There also exists a term called $\gamma$ which is deliberately used to capture the sensitivity of resistance to the filament constriction process. Higher values of $\gamma$ indicate that filament geometry is highly sensitive to small geometric changes in the filament where even minor variations like migration of a few Ta cations would lead to significant changes in the resistance of the device; in contrast, for lower values of $\gamma$, the resistance of the filament is insensitive to changes in filament geometry where even major events like set and reset processes are not characterized properly. Hence, keeping these two limiting conditions in mind, we should carefully tune the coefficient of $\gamma$ within a physical range that would yield accurate and reproducible IV characteristics where the log-log plot of the same would be capable of highlighting the electron conduction mechanism of the device. Overall, this term helps capture subtle atomistic features associated with filament constriction and bridges it with the measurable electrical characteristics, thereby resulting in a comprehensive filament structure and IV characteristics. During the set process (V $>$ V$_{set}$, V˙ $>$ 0), our MD simulations indicate that the filament growth is abruptly accelerated with increasing field beyond the threshold required to begin the set process. This accelerated filament growth is purely because of Ta cation diffusion driven by electric field bias. Hence, for this process we model the growth of the filament through a first order kinetic equation shown in Eq.~\ref{eq:set_kinetics}.

\begin{equation}
\label{eq:set_kinetics}
\frac{dx}{dV}
= K_{\mathrm{set}} \frac{1 - x}{V},
\qquad V \ge V_{\mathrm{set}}, \quad \dot{V} > 0
\end{equation}

This particular functional form was chosen solely because if we closely observe the MD trajectories during the set process, we see that filament growth scales with 1-$x$. That is because growth slows down as the filament formation approaches its completion. Another important aspect of the kinetic equation is that we want to preserve the value of $x$ between 0 and 1. To achieve this, we assume that the filament growth rate varies inversely with the applied field bias. This assumption arises from making direct observations from the MD simulations which show that when the external electric field value exceeds the threshold required to initiate the set process, the newly formed filament grows slowly, indicating that the growth process is approaching its completion. On integrating Eq~\ref{eq:set_kinetics}, we get Eq.~\ref{eq:set_x} which defines the functional form of $x_{set}$ which would approach unity as we continuously increase the field until we reach the extreme end of the applied field. This is consistent with MD simulations where once the external electric field reaches its maximum value, the number of Ta cations forming the filament also approaches its maximum, therefore showcasing filament stabilization.

\begin{equation}
\label{eq:set_x}
x_{\mathrm{SET}}(V) =
1 - (1 - x_0)\left( \frac{V_{\mathrm{set}}}{V} \right)^{K_{\mathrm{set}}}
\end{equation}

\begin{equation}
\label{eq:x_min}
x_0 = \min \left( \frac{N_{\mathrm{Ta}}(V)}{N_{\mathrm{Ta,max}}} \right)
\end{equation}

During the reset process (V $>$ V$_{reset}$, V˙$<$ 0), the filament first undergoes constriction before it completely ruptures, breaking the conduction pathway between the two electrodes. For this process, the kinetic rate equation is described by Eq.~\ref{eq:reset_kinetics}. This kinetic equation represents a first order decay which, when integrated, would yield Eq.~\ref{eq:x_reset} where $x$ during reset would exponentially decay with increasing electric field beyond the threshold required to initiate the onset of reset. This characteristic can also be validated from MD where going beyond the threshold field causes an abrupt rupture in the filament.

\begin{equation}
\label{eq:reset_kinetics}
\frac{dx}{dV}
= -K_{\mathrm{reset}} \, x,
\qquad |V| \ge V_{\mathrm{reset}}, \quad \dot{V} < 0
\end{equation}

\begin{equation}
\label{eq:x_reset}
x_{\mathrm{RESET}}(V) =
x_{\max} \exp \left[ -K_{\mathrm{reset}} \left( |V| - V_{\mathrm{reset}} \right) \right]
\end{equation}

\begin{equation}
\label{eq:x_max}
x_{\max} = 1
\end{equation}

For the regions that lie outside the kinetic modeling framework of set and reset processes, we obtain the value of x by using Eq-5. For current calculation, we have two terms. One term is the current that is being obtained due to the evolution of the conductive filament over a cycle of operation and the other term is the leakage current that is evident in almost all memristive systems. The current obtained by filament formation and rupture process is modeled by a simple ohmic relation given in Eq.~\ref{eq:filament_current}, whereas the leakage current is modeled by a hyperbolic function of applied voltage and a fitting constant $\alpha$ as shown in Eq.~\ref{eq:leakage_current}. The hyperbolic function for leakage current arises from the combination of Poole-Frenkel emission~\cite{Schroeder2015PooleFrenkel} and trap-assisted tunneling~\cite{Wu1999TrapAssistedTunneling}. Both of these mechanisms have proven their capability to model leakage current in bulk oxides like HfO$_2$ and Ta$_2$O$_5$.

\begin{equation}
\label{eq:filament_current}
I_{\mathrm{filament}} = \frac{V}{R(x)}
\end{equation}

\begin{equation}
\label{eq:leakage_current}
I_{\mathrm{leakage}} = I_{\mathrm{leak}} \sinh (\alpha V)
\end{equation}

The final total current is shown in Eq.~\ref{eq:final_current} which is just a simple summation of filament and leakage currents.

\begin{equation}
\label{eq:final_current}
I = I_{\mathrm{filament}} + I_{\mathrm{leakage}}
\end{equation}

All parameters employed in modeling the IV characteristics are listed in Table~\ref{tab:parameters}. The scaling factors used to convert the electric field values applied in the MD simulations to the corresponding voltages needed for analytical modeling are well summarized in Table. 1 of supplementary information.
\vspace{0.5em}
\begin{table}[h]
\centering
\caption{Device parameters used in the analytical modeling of the Ta–HfO$_2$–Pt memristive device.}
\label{tab:parameters}
\begin{tabular}{|c|c|c|c|c|c|}
\hline
\textbf{$R_{\mathrm{on}}$ ($\Omega$)} & 
\textbf{$\gamma$} & 
\textbf{$I_{\mathrm{leak}}$ (A)} & 
\textbf{$\alpha$ (V$^{-1}$)} & 
\textbf{$K_{\mathrm{set}}$} & 
\textbf{$K_{\mathrm{reset}}$} \\ \hline
$1 \times 10^{3}$ & 7.5 & $2 \times 10^{-5}$ & 2.0 & 1 & 0.65 \\ \hline
\end{tabular}
\end{table}

\section{Results and Discussions}

\subsection{Electric-Field-Induced Interfacial Reconstruction and Filament Nucleation}
\begin{figure}[H]
    \centering
    \includegraphics[width=1\textwidth]{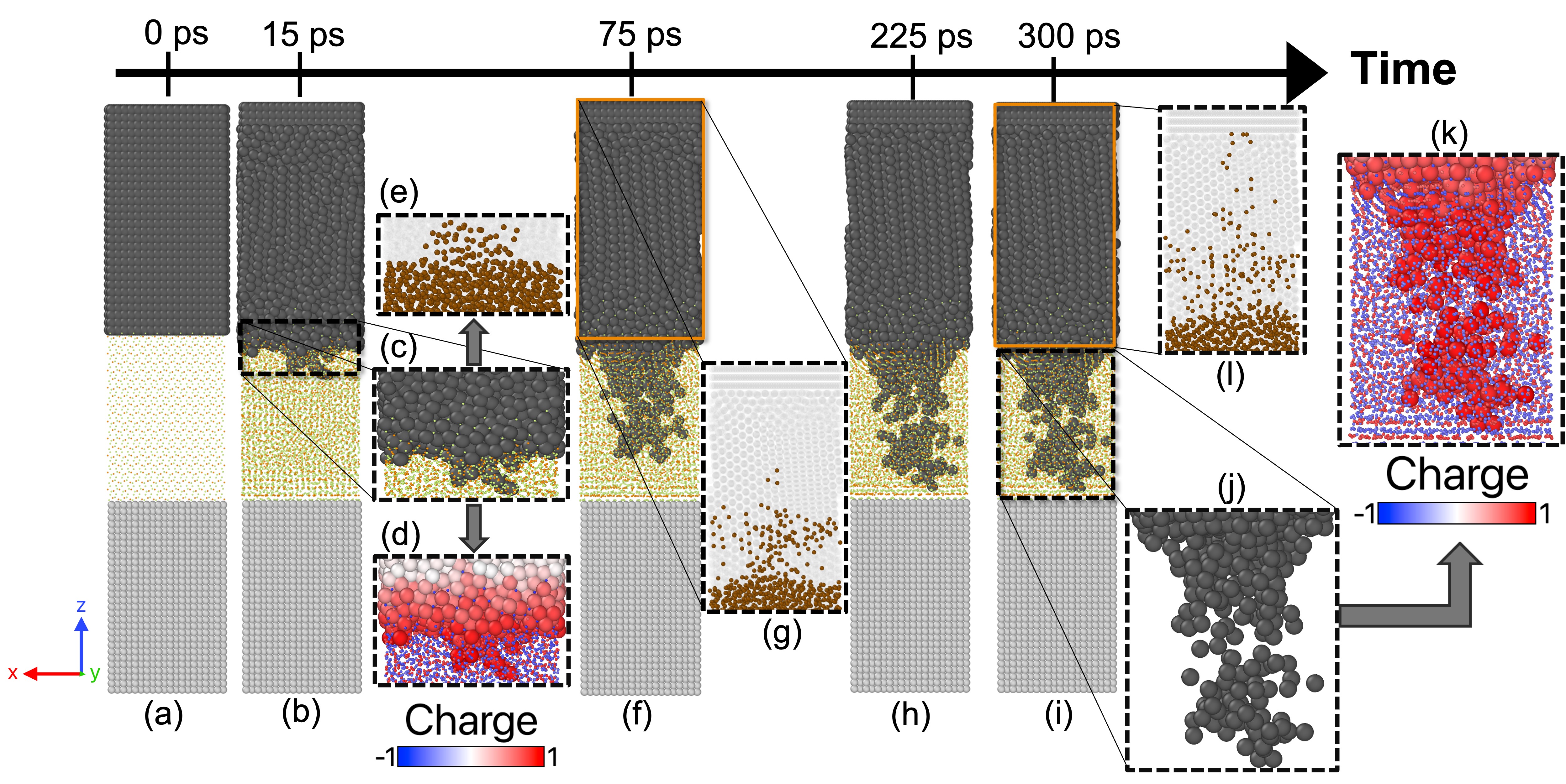}
    \caption{ \textbf{A detailed step by step filament formation process of Ta-HfO$_2$-Pt memristor}. (a) The initial state of the device before an external electric field was applied, (b) shows the state of the system when a non-stoichiometric layer of TaO$_x$ at the Ta-HfO$_2$ interface is formed, (c) The magnified snapshot of the non-stoichiometric layer of TaO$_x$, (d) The charge distribution of the non-stoichiometric layer of TaO$_x$, (e) highlights the diffusion of oxygen anions at the metal-semiconductor interface which is governing the formation of TaO$_x$, (f) shows the state of the system when Ta cations begin to diffuse hence marking the onset of the filament formation process (g) points the position of oxygen anions which begin to spread across the top electrode. (h) An intermediate state of the device before final stable filament formation. (i) Final filament formed composed of Ta cations, where (j) points towards a magnified snapshot of the filament, (k) maps charges onto the final configuration of the system highlighting the ionic nature of the filament and (l) shows the final state of oxygen anions which are distributed across the top electrode.}
    \label{fig:framework}
\end{figure}

We start by performing static electric field analysis on Ta-HfO$_2$-Pt to understand the steps governing the filament formation. The electric field applied in this particular case was fixed at 1 V/Å and was applied along the negative z-direction. The bottom Pt electrode was fixed to mimic the ground state nature of this electrode. As soon as the field was applied, as seen in Figure~\ref{fig:framework}(b,c,d,e), the Ta cations of the top electrode diffused and formed a non-stoichiometric layer of TaO$_x$ at the interface within the first 15 ps of the simulation. The formation of the non-stoichiometric layer can be confirmed by mapping atomic charge onto the structure where clearly the Ta cations attain a high positive charge with oxygen anions attaining a negative charge, but the rest of the electrode is still charge neutral, confirming its metallic nature. The reason behind the quick expedited nature of the non-stoichiometric layer formation is that the Ta cations diffuse very rapidly owing to their very low activation barrier of 0.04 eV~\cite{Davies1965Migration, Whitton1968IonicMobilities} which is required for them to start diffusing from their initial positions in the crystal structure. The Ta ions present in the interfacial layer of TaO$_x$ are highly diffusible in nature, due to which they start forming a conductive bridge with the Pt electrode as seen in Figure~\ref{fig:framework}(f,g,h,j,k). The region of HfO$_2$ surrounding the filament becomes amorphous in nature. These observations match with previous STEM and EELS spectrum findings. For a memristor to behave like an ECM cell, the filament formed should be metallic in nature; that is, when the metal ions diffuse and form a bridge, they lose their charge to attain charge neutrality, and hence the filament is just composed of neutral metal atoms. On the other hand, in a VCM cell we hardly observe any metal ion diffusion. However, for this particular memristive device, the final filament is neither composed of only oxygen vacancies nor do the Ta ions lose their charge to form a metallic bridge; rather, the filament formed is indeed an amorphous solid solute of Ta(O). This can be verified by mapping the charges onto individual atoms where clearly from fig.2(i) we can observe that the Ta filament formed is still in its ionic phase, which is surrounded by randomly distributed oxygen anions. A filament composed of Ta ions is not the only governing switching mechanism. Since the diffusion barriers for both Ta and O are very low, the oxygen anions present in HfO$_2$ also respond to the external electric field, thus migrating towards the top electrode. Initially, the ratio of O/Hf was 2, but since the oxygen anions diffused towards the top electrode, this ratio rapidly decreases to 1.5 after the filament formation, which can be observed structurally from Figure~\ref{fig:framework}(l). This oxygen deficient region formed during the application of an external field is consistent with previously reported experimental findings~\cite{Jiang2016Sub10nmTa}; hence, this particular memristive device can be put into a group of modified VCM cells where both oxygen deficient regions and metal cation rich regions switch the device from HRS to LRS. Moving forward, we would be performing a full cycle of operation on this memristive device in order to understand the atomistic details associated with LRS, HRS, and other intermediate states which are important to achieve sustained memory operations using a cross-bar array of memristors.

\subsection{Full-Cycle Set/Reset Dynamics and Filament Rupture}
\begin{figure}[H]
    \centering
    \includegraphics[width=1.0\textwidth]{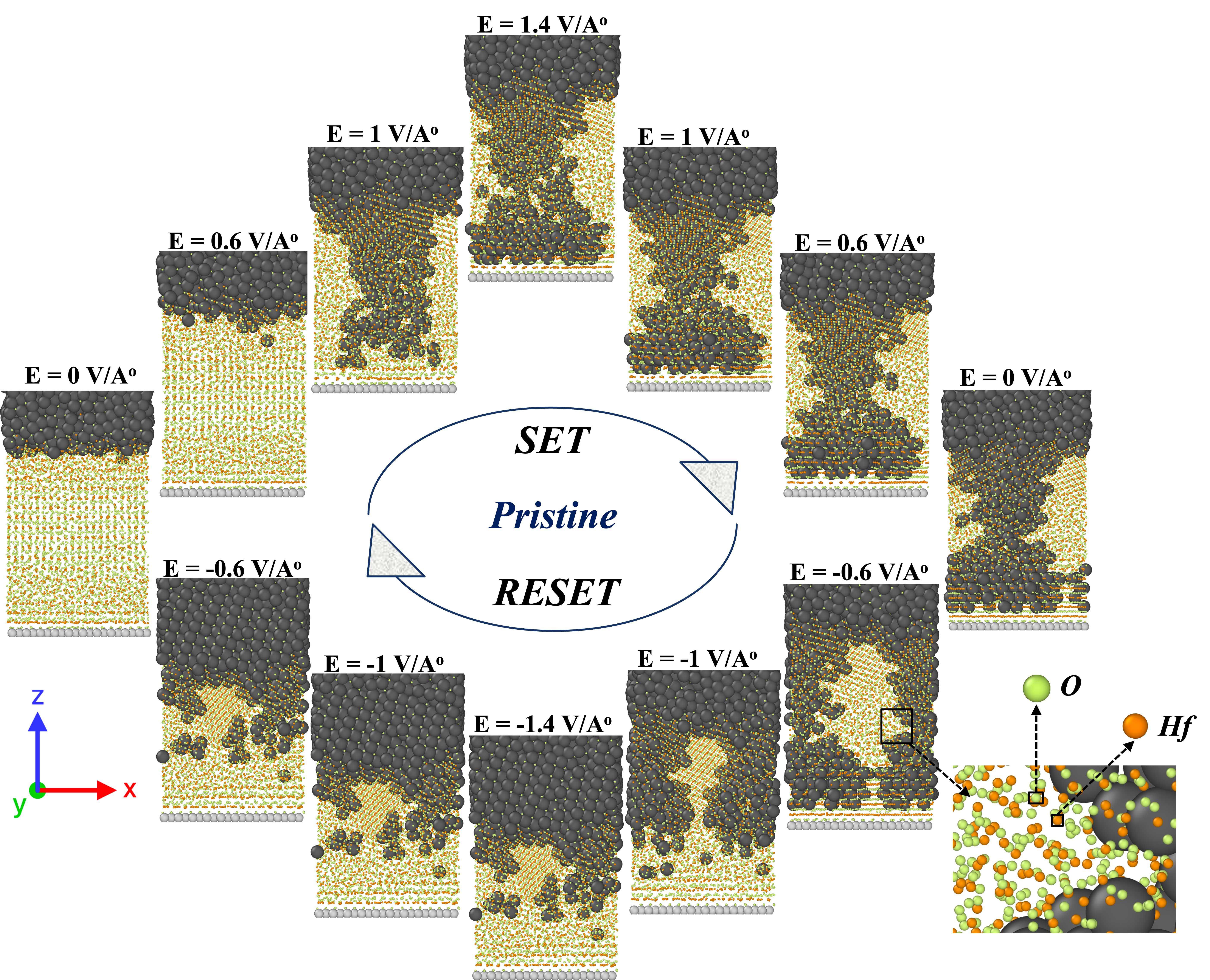}
    \caption{\textbf{Full-Cycle switching behavior of pristine device}. After the completion of the set process a filament composed of Ta cations is formed between the electrodes causing the device to switch to LRS. The reset process of the device occurs in a step-wise manner with first filament constriction observed at E = -1 and -1.2 V/Å followed by a full filament rupture at -1.4 V/Å. After complete filament rupture the device switches back to HRS with the state being stable owing to minimal changes in the spatial distribution of atoms.}
    \label{fig:full-cycle-pristine}
\end{figure}

For a full cycle of operation on the Ta-HfO$_2$-Pt memristive system, we observe an immediate formation of the non-stoichiometric layer of TaO$_x$ at 0 V/Å. As observed from Figure~\ref{fig:full-cycle-pristine}, Ta cations would not diffuse to form a conductive bridge until the external electric field reaches a value of 0.8 V/Å. At this field, a half filament is formed, and as soon as the field reaches a value of 1 V/Å, a full conductive bridge composed of Ta cations is formed. As the field increases up to 1.4 V/Å, the number of Ta cations forming the filament continuously increases but then remains constant until the field cycle reaches the reset threshold. The onset of the reset process is characterized by a reduction in the number of Ta cations forming the filament. As observed from fig.4(b), the reset process begins at an electric field value of -1 V/Å where the filament starts to constrict in size. The filament constricts further at -1.2 V/Å and fully ruptures at -1.4 V/Å. Structurally, we can observe two more stable resistance states arising from the reset process: one resistance state at -1 V/Å and the other at -1.2 V/Å. These two states are distinguishable from our regular LRS and HRS structures. This shows that Ta-HfO$_2$-Pt memristive systems are capable of projecting multi-resistance states in a single cycle, which is key to realizing ultra-fast non-volatile memories in the form of oxide based memristors. To establish a proper correlation between the structures obtained from MD and the final electrical characteristics of the device, we made use of an analytical model (see Methods section) to obtain the IV characteristics of the MD structures. In this model, we make use of the number of Ta cations forming the filament and the number of Ta cations left after the filament is fully ruptured. 

\subsection{Analytical I–V model and structure–property correlation}

\begin{figure}[H]
    \centering
    \includegraphics[width=1\textwidth]{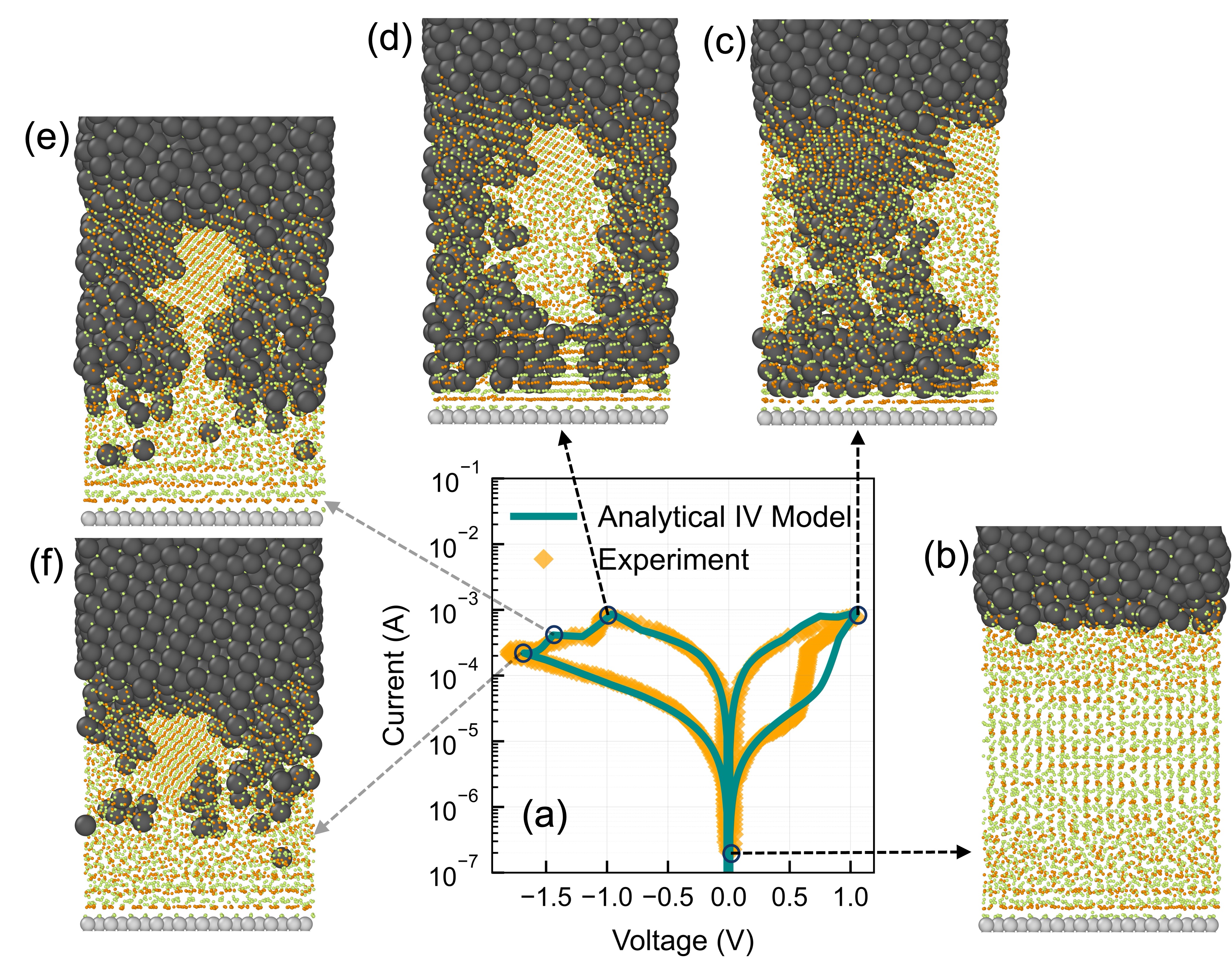}
    \caption{\textbf{Correlation between IV characteristics and filament evolution in pristine Ta-HfO$_2$-Pt device}.(a) shows the IV characteristics of pristine Ta-HfO$_2$-Pt device. (b) Depicts the initial state of the device at 0 V, (c) highlights the LRS which structurally corresponds to a full filament composed of Ta cations. (d),(e) points towards the stepwise filament constriction before (f) full filament rupture occurs.}
    \label{fig:IV_snaps}
\end{figure}

The IV characteristics obtained can be seen in Figure~\ref{fig:IV_snaps}(a,c), where clearly in the set cycle we only have LRS as our stable state to be used for memory applications. We observe 2 intermediate resistance states during the reset process at -1 and -1.2 V which is shown in Figure~\ref{fig:IV_snaps}(a,d,e), which is consistent with what was observed structurally. The analytical model proposed is indeed consistent with experimental findings~\cite{Jiang2016Sub10nmTa}, which showcases the robustness of this model. This model is computationally cheap and avoids the need for expensive models like NEGF based Quantum-transport method or Tight-Binding approach. However, this model is limited to filamentary memristors and requires an appropriate charge-based force field for the specific material system. To test the predictive capability of the model, we proceed with modeling the same Ta-HfO$_2$-Pt memristive system, but this time we introduce different concentrations of oxygen vacancies in the initial structure. We also test the model's capability by varying the concentration of oxygen vacancies present in our initial HfO$_2$ structure.

\subsection{Validation of the Analytical Model for IV Characteristics and Electron Conduction Mechanisms}
In our developed analytical model, the evolution of device resistance is correlated with filament geometry through an exponential term. In this functional form, we include a coefficient $\gamma$ which is not introduced as a simple fitting parameter but which naturally arises from the quantum transport of electrons with a barrier to their flow. During filament constriction, the dominant electron conduction changes from metallic, ohmic conduction to barrier limited tunneling, where the reduction in the number of Ta cations leads to the formation of atomic scale gaps between the filament and the bottom electrode. In addition to this, the cross-sectional area of the conductive bridge also decreases. When these factors are coupled, we observe an increase in tunneling barrier height and width; therefore, electrons experience higher hindrance to tunnel. Since tunneling probability has an exponential correlation with the barrier characteristics, even small changes in the geometric structure of the filament would cause large changes in device resistance. Upon complete filament rupture, the device resistance increases exponentially, causing it to switch from LRS to HRS. The extent to which structural changes are translated into an electrical response is governed by the sensitivity coefficient $\gamma$. This parameter helps us control how strongly resistance would respond to changes in filament geometry. For capturing all the necessary processes, the value of $\gamma$ should range from 5 to 8, where the model is capable of accurately capturing filament growth, constriction, and rupture. $\gamma$ values lying outside the mentioned range would cause a large divergence from physical reality. 

\vspace{0.5em}

The kinetic equations governing the SET and RESET processes are derived directly from the field driven diffusion of Ta cations when MD simulations were performed on the memristive system. During the SET process, the filament growth at first increases rapidly before it slows down and drops to almost negligible values, highlighting that the filament is fully grown and is now stable. Hence, in this entire process we observe a need for a kinetic equation that can model the slow growth of the filament at electric fields greater than the threshold required to initiate the SET process. Hence, we adopt a first order kinetic equation where the growth rate of the filament is directly proportional to (1-$x$). The first reason to use such a form is to ensure that the value of $x$ is within the range of (0,1). Moreover, in the kinetic equation we consider the growth rate to be inversely proportional to the applied bias voltage. This means that as the filament continues to grow, the already existing Ta cations hinder the migration of Ta cations from the top Ta electrode. Conversely, the RESET process is characterized by filament constriction before full filament rupture. Such a type of behavior depicts a decay process. Hence, for RESET we assume that the rate of filament rupture is directly proportional to the state variable $x$. Such proportionality yields a first-order exponential decay in the filament structure. However, such kinetics help us model gradual constriction and full filament rupture with utmost accuracy. 

\vspace{0.5em}

\begin{figure}[H]
    \centering
    \includegraphics[width=1\textwidth]{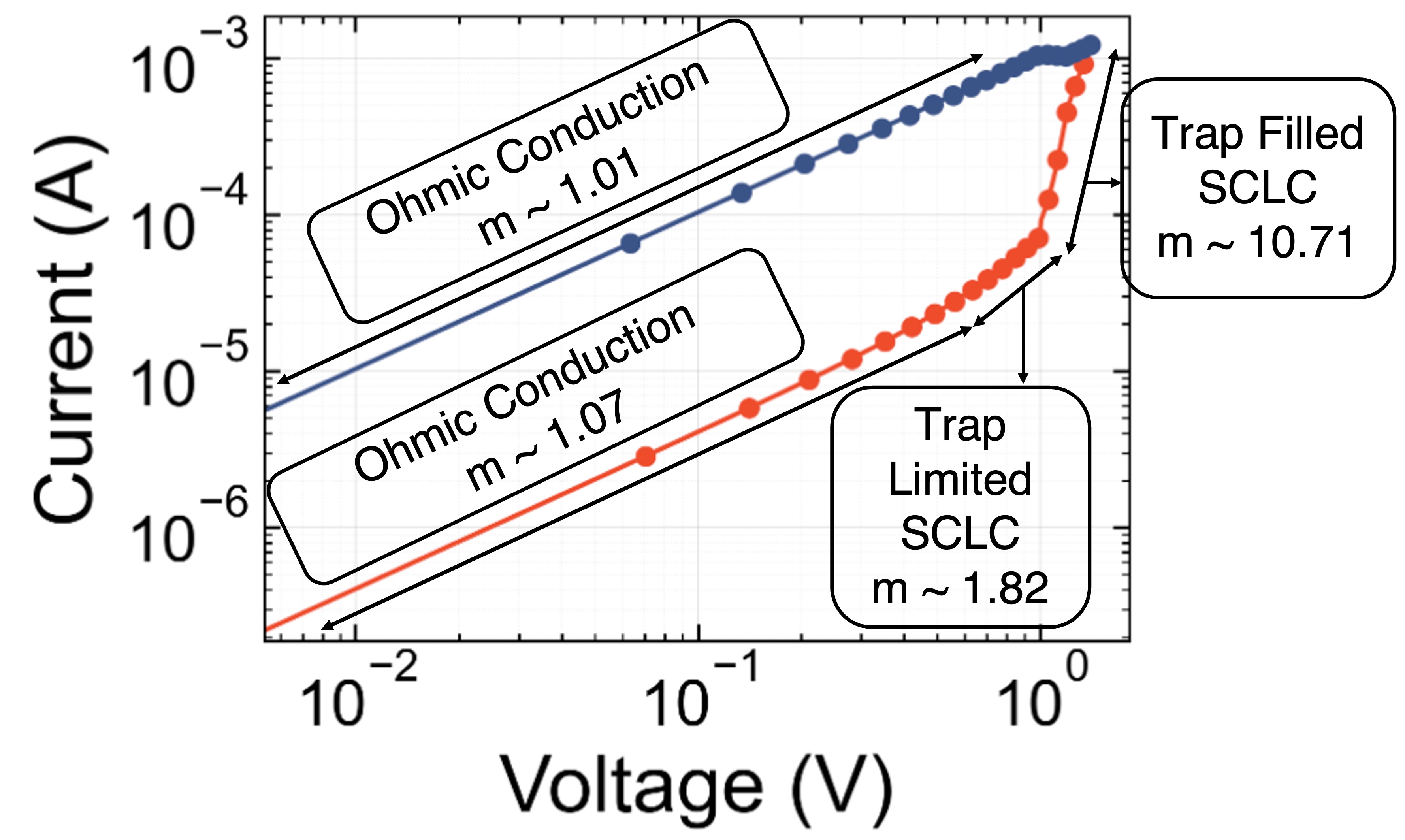}
    \caption{\textbf{Exploring electron conduction mechanism governing the resistive switching of the device}. The log-log plot of current and applied bias voltage where before the set process the dominant mechanism is ohmic, during the transient state of SET process its trap limited SCLC, during the set process its trap filled SCLC and during the second half of positive voltage cycle the dominant mechanism again switches back to ohmic.}
    \label{fig:electron_conduction}
\end{figure}

IV characteristics provide critical insights into the dominant electron conduction mechanisms at different stages of full cycle field operation. To identify these mechanisms, we obtain a log-log plot of the IV characteristic and then analyze the slope of the curve where the magnitude of the slope helps determine the prevailing transport mechanism in different voltage regimes. As shown in Fig.~\ref{fig:electron_conduction}, prior to the set process, the device predominantly exhibits ohmic conduction. 

\vspace{0.5em}

In this region, the current would scale linearly with the applied bias voltage, indicating that the charge transport is still being governed by intrinsic material properties; that is, the conduction seen at this state of the device can be mainly attributed to thermally generated carriers with negligible presence of external charge carriers. This finding is consistent with MD simulations where in this regime the onset of filament formation has not begun. In the voltage region close to the set voltage, a small increase in the slope is observed. The slope value for this region is approximately 1.82, which indicates that in this region the conduction of electrons is governed by trap-limited space-charge-limited current (SCLC)~\cite{conduction}. This behavior corresponds to the onset of filament growth, during which, as the Ta cations migrate towards the bottom electrode, more and more oxygen vacancies are created. These vacancies introduce localized trap states in the band structure of HfO$_2$. The electrons generated by the filament are captured by these trap states; hence, these traps are impeding smooth electron transport. As a result, the current no longer scales linearly with voltage. For voltages that exceed the set voltage, a sharp rise in current is observed, which is a typical characteristic of abrupt resistive switching. In the corresponding log-log IV plot, we obtain a slope of 10.71 for this region. This indicates that the electron conduction mechanism has transitioned to trap-filled SCLC~\cite{conduction}. In this regime, the previously partially filled trap states are now fully occupied. Consequently, additional electrons generated by a fully formed conductive filament are no longer captured by these traps. With these traps fully saturated with electrons, we can now observe charge transport almost free from any external barriers. This behavior reflects the establishment of a highly conductive filament based pathway between the two electrodes, enabling efficient electron transport across the device. During the second half of the voltage cycle, we observed that the conduction mechanism switches back to ohmic conduction. This means that the conductive filament formed during the SET process has now achieved stability. Since the trap states are already filled and the filament is fully developed, the charge transport is no longer limited by traps or space-charge effects. Instead, the device now showcases steady charge transport dominated by saturated external charge carriers that were generated by the filament in the regime of trap-filled SCLC.

\subsection{Effect of Oxygen Vacancy Concentration on Filament Formation and Rupture}

\begin{figure}[H]
    \centering
    \includegraphics[width=1\textwidth]{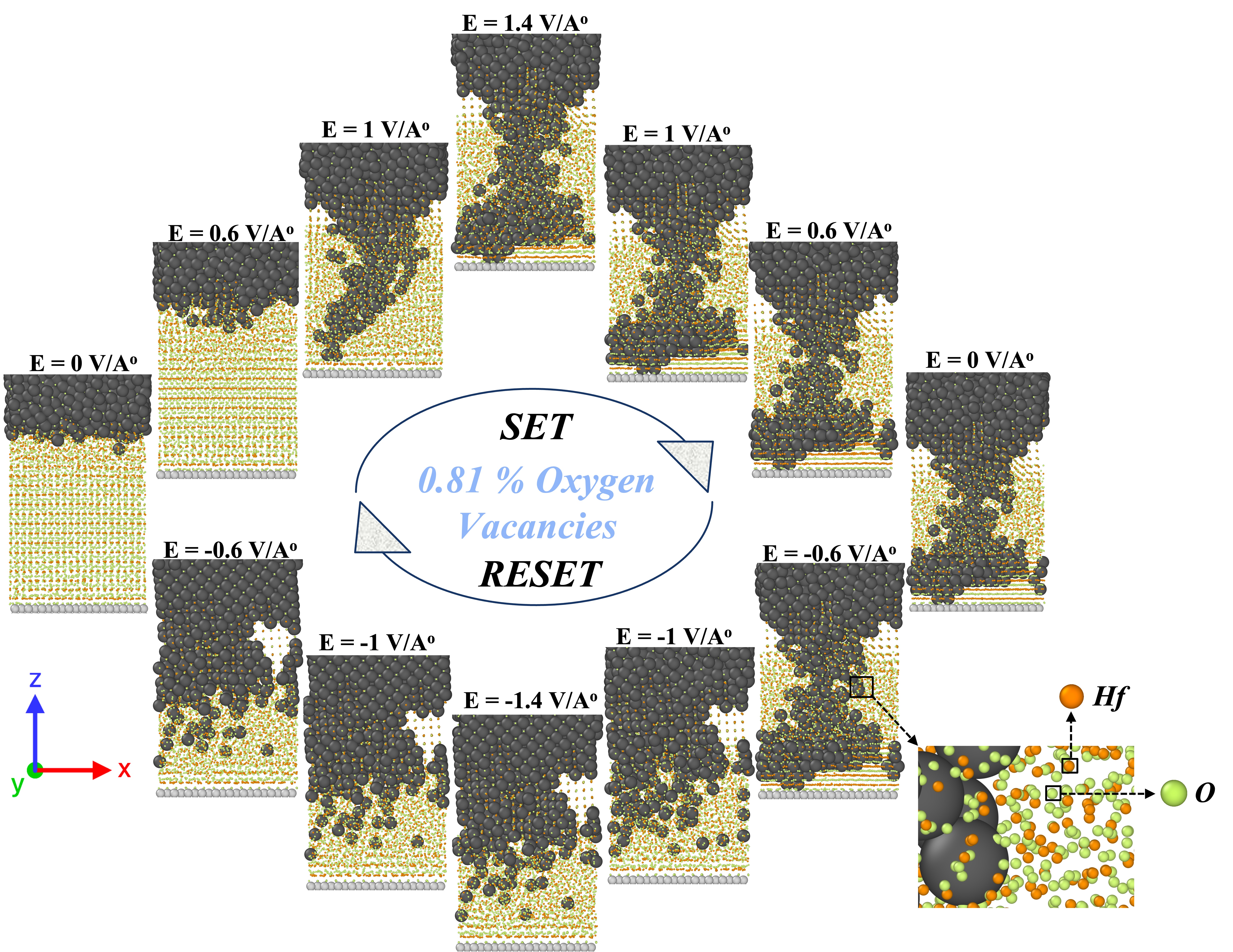}
    \caption{\textbf{Full-Cycle switching behavior of the device with 0.81\% oxygen vacancies}. This device exhibits a straightforward response to an electric field cycle with both set and reset processes occurring sharply at 1 and -1 V/Å, respectively.}
    \label{fig:0.81_percent_oxygen_vacancies_full_cycle}
\end{figure}

\begin{figure}[H]
    \centering
    \includegraphics[width=1\textwidth]{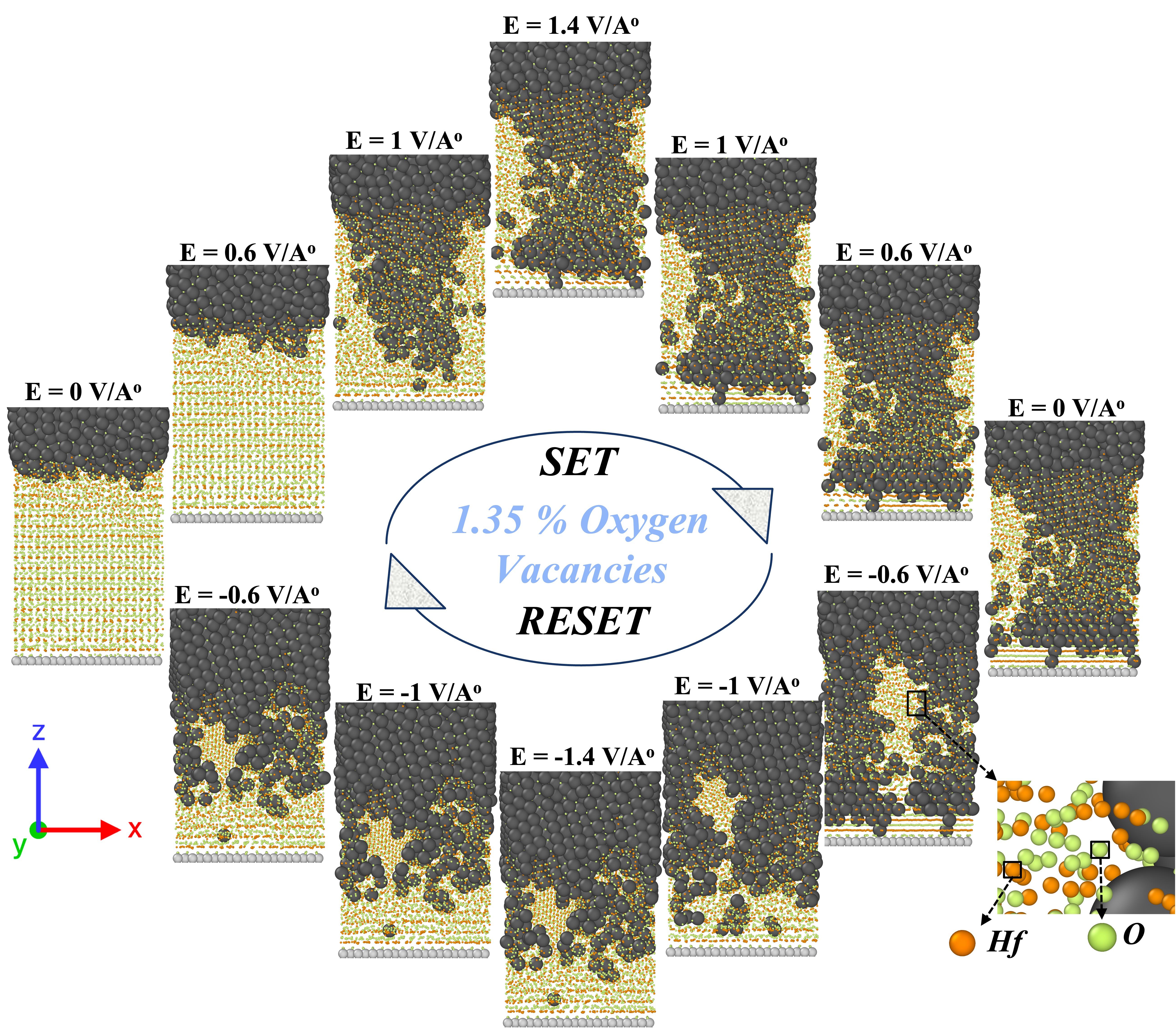}
    \caption{\textbf{Full-Cycle switching behavior of the device with 1.35\% oxygen vacancies}. The completion of the set process is marked by the formation of a filament composed of Ta cations. Similar to the pristine device, a stepwise reset process is observed with filament constriction taking place at -1 V/Å followed by a full filament rupture at -1.4 V/Å.}
    \label{fig:1.35_percent_oxygen_vacancies_full_cycle}
\end{figure}

To study the effect of different oxygen vacancy concentrations on the switching characteristics of the device, an initial oxygen vacancy concentration of 0.81\% was introduced to the HfO$_2$ structure. This device was then subjected to an external electric field cycle. Up to an electric field value of 0.6 V/Å, only a non-stoichiometric layer of TaO$_x$ was formed (as seen in Figure~\ref{fig:0.81_percent_oxygen_vacancies_full_cycle}), which is consistent with previous findings associated with the pristine Ta-HfO$_2$-Pt memristive device. An electric field of 0.8 V/Å marked the onset of Ta cation diffusion, and these cations continued to diffuse until the electric field reached a value of 1.4 V/Å, which marks the end of the filament formation process. During the reset process, we observe a striking difference between the two devices (pristine and defective device with 0.81\% oxygen vacancies). In the pristine device, we had observed a step by step filament constriction which had yielded two intermediate memory states before the filament was completely ruptured to switch back to its HRS. However, in the case of a device with initial oxygen vacancies, the filament was completely ruptured at -1 V/Å (as observed in Figure~\ref{fig:0.81_percent_oxygen_vacancies_full_cycle}); therefore, this particular device configuration lacks the ability to be utilized in memory devices where multi-level resistance states are a must. Another interesting fact to note is that the filament shape and size are different for both cases; in the device with initial defects, the final filament shape resembled an hourglass. Owing to its near perfect filament shape and size, it was easier to rupture the filament at lower fields, hence giving us more control over the filament formation and dissolution process. In addition to 0.81\% oxygen vacancies, we also explore the process of filament formation and rupture for 1.35\% oxygen vacancies in the initial HfO$_2$ crystal structure in order to have a better understanding of the role of defects in governing the switching characteristics of the device. For the device with 1.35\% oxygen vacancies, we observed an onset of Ta cation diffusion at an electric field value of 0.8 V/Å, which is similar when compared to what was observed for the device with 0.81\% oxygen vacancies and pristine device configurations. A full filament is formed at 1.4 V/Å. As we observe from Figure~\ref{fig:1.35_percent_oxygen_vacancies_full_cycle}, the shape of the filament formed resembles a near perfect hourglass, which was earlier seen in the device which had 0.81\% of oxygen vacancies in its initial structure. Figure~\ref{fig:1.35_percent_oxygen_vacancies_full_cycle} further shows that the reset process is characterized by a stepwise filament constriction before leading towards filament rupture. These observations are consistent with the pristine Ta-HfO$_2$-Pt device.

\begin{figure}[H]
    \centering
    \includegraphics[width=1\textwidth]{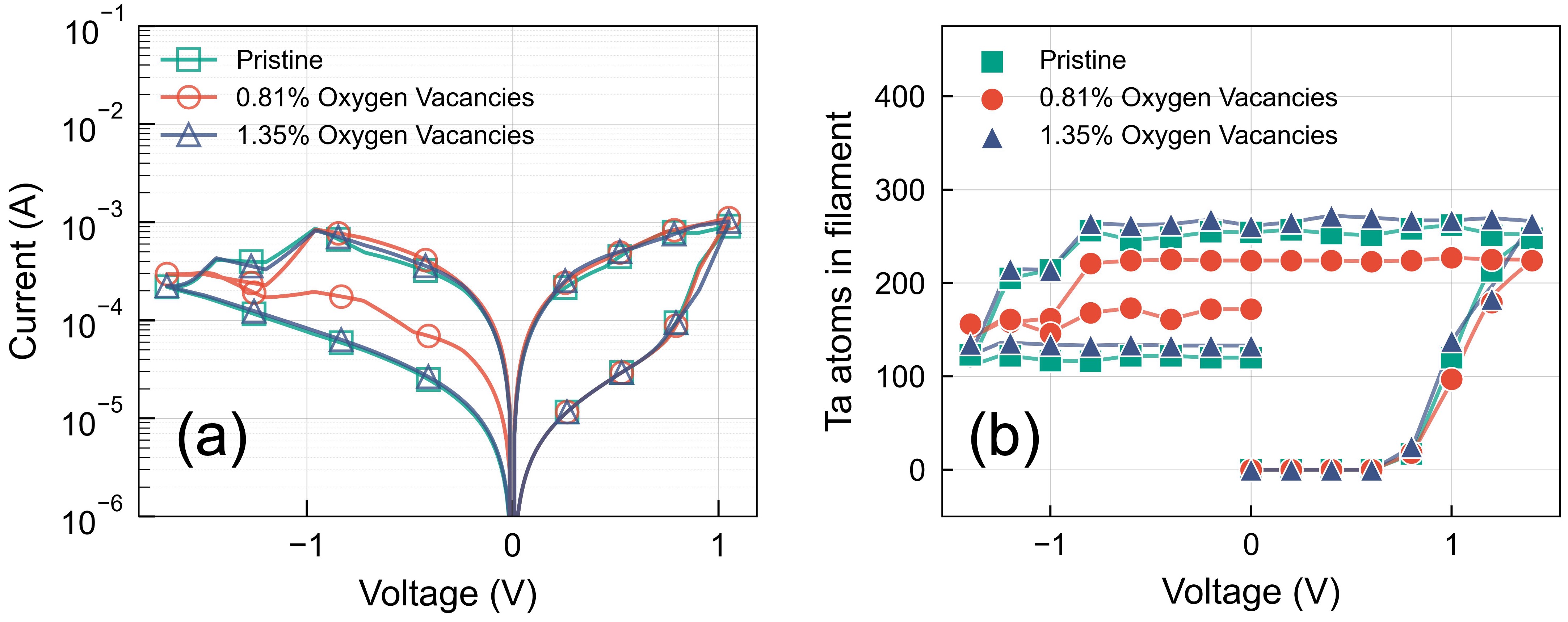}
    \caption{\textbf{Electric field driven conductive filament dynamics}. (a) IV Characteristics obtained from the developed analytical model that makes use of the trajectories of Ta cations forming the conductive bridge between the two electrodes for all three cases: pristine, 0.81\% oxygen vacancies and 1.35\% oxygen vacancies,(b) shows the electric field dependent dynamics of Ta cations in the filament region for all the three device configurations}
    \label{fig:all_device_cases_IV}
\end{figure}

From Figure~\ref{fig:all_device_cases_IV}(a), we observe that the SET cycle for all the devices perfectly overlaps. In contrast, the RESET cycle exhibited noticeable variation. For the pristine device and the device with 1.35\% oxygen vacancies, the resistance increases in a gradual stepwise manner, which can be directly correlated with the stepwise filament constriction before full filament rupture that is seen in these devices. However, for the device with 0.81\% oxygen vacancies, the RESET transition is abrupt, consistent with the atomistic picture in which the conductive filament undergoes sudden rupture at the RESET voltage without intermediate stages of constriction. Figure~\ref{fig:all_device_cases_IV}(b) further shows that we have an interesting observation where the number of Ta cations forming the conductive bridge in the memristive device with 0.81\% oxygen vacancies is slightly lower compared to the pristine device, but the same number was slightly higher for the device with 1.35\% oxygen vacancies. These results indicate that the initial presence of defects influences both the morphology of the filament during the SET process and the residual state of the device after reset. However, a fundamental theory that can quantitatively link the initial defect network with the final filament geometry is still lacking. This gap highlights a major long standing issue of device to device variability in memristors. In practical crossbar arrays, such variations in defect concentration and spatial distribution across different devices will lead to inconsistencies in the overall switching behavior of the device. 

\subsection{Electric Field Pulse Operations on Pristine and Defective Memristors}
\begin{figure}[H]
    \centering
    \includegraphics[width=1\textwidth]{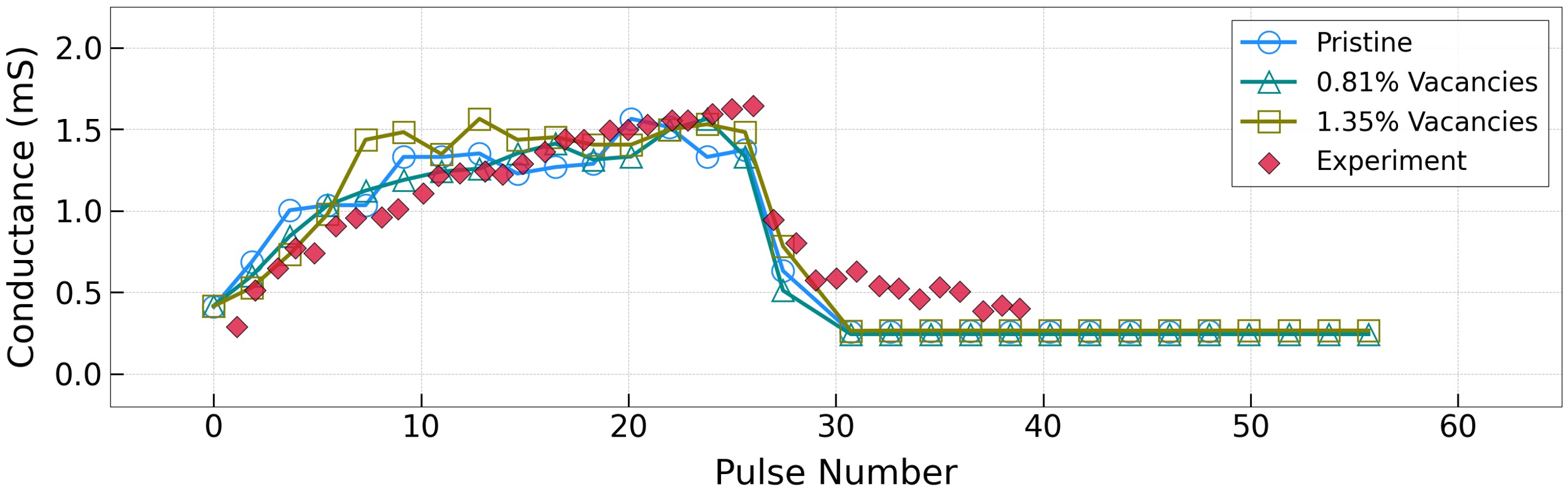}
    \caption{\textbf{Neuromorphic behavior of the device under a field pulse operation}. is evaluated for all different configurations which are: (i) Pristine, (ii) 0.81\% and (iii) 1.35 \% . The final output of this study is then compared to previously reported experimental data~\cite{Jiang2016Sub10nmTa} for evaluating the robustness and predictive capability of the developed analytical model.}
    \label{fig:conductance-pulse-number}
\end{figure}

An important application of memristors is their use in the construction of a neuromorphic architecture. In this application, memristors are used to duplicate the functionality of a biological neuron, which enables the replication of a computing architecture that works like a human brain. To evaluate the performance of memristive devices for such applications, it is important to perform electric field based pulse operation on all three cases of our device, i.e., pristine, 0.81\% oxygen vacancies, and 1.35\% oxygen vacancies. After applying electrical pulses, we obtain Figure~\ref{fig:conductance-pulse-number} which shows a plot of conductance versus pulse number for all three cases of the device, which are then compared with previously reported experimental data~\cite{Jiang2016Sub10nmTa}. Among the three cases, the device with 0.81\% oxygen vacancies performed the best, where the conductance almost varied linearly with respect to the pulse number, which indicates that the filament growth in this situation was stable and reliable. This device also matched with experimental findings~\cite{Jiang2016Sub10nmTa}. However, the results of the device with 1.35\% oxygen vacancies deviated significantly from the previous case. In this device, we observe a non-linear response to the positive pulse cycle, which arises due to a sudden rise in the population of Ta cations forming the filament. However, the filament formed in this device is not stable. The Ta cations remain mobile and constantly move back and forth between the electrode and the oxide layer. This observation suggests that the Ta cations are still undergoing restructuring before forming a filament that is both structurally and thermodynamically stable. These observations indicate that oxygen vacancy concentration plays a critical role in pulsed cycle operations as well. In one case, a linear response with stable and continuous filament formation is observed whose conductance states are suited to mimic the synaptic weights over multiple pulsed cycle operations; on the other hand, the device with higher initial vacancy concentration resulted in a stochastic response to the same pulsed cycle that induced high fluctuations in the conductance values, making it undesirable for neuromorphic applications. In addition to the device with 1.35\% oxygen vacancies, the pristine device also showed deviations from the previously reported experimental data~\cite{Jiang2016Sub10nmTa}. However, the magnitude of stochasticity in the pristine device was comparatively lower, suggesting that the filament formed in the pristine device is more controlled compared to the previous case, even though a certain degree of dynamic restructuring of the filament still exists. All these results highlight the stochastic nature of device performance when defects are introduced to the device. This nature of the device is acting as a significant barrier to its commercialization. One way of controlling the stochastic behavior of filament formation and rupture is to control the concentration and spatial distribution of defects in the initial structure. With defect engineering of the device, we can expect a deterministic kinetic pathway for filament formation and the dissolution process, thereby giving us more control over the device when exposed to different electrical stimuli. Such strategies will also help in the fabrication of reproducible devices with reduced variability and enhanced endurance and retention characteristics.

\section{Conclusion}
Our work provides an atomistically resolved understanding of resistive switching in Ta--HfO$_2$--Pt memristive devices by demonstrating that the switching process is governed by the coupled formation and rupture of a hybrid Ta-cation-rich, oxygen-deficient conductive filament, rather than by a purely oxygen-vacancy-based pathway. Molecular dynamics simulations with dynamic charge transfer reveal the sequence of electric-field-driven interfacial reconstruction, TaO$_x$ formation, Ta cation migration, and oxygen redistribution within HfO$_2$, thereby establishing a physically grounded picture of filament nucleation, growth, constriction, and rupture. By linking filament evolution to an analytical current--voltage framework, we connect atomistic structural changes to experimentally relevant electrical characteristics and capture key features such as intermediate resistance states and variability arising from the initial defect landscape. 

Our framework enables us to obtain IV characteristics directly from MD trajectories, hence bridging the gap between the atomistic switching mechanism and macroscopic electrical properties. Since this framework is capable of establishing a coherent relation between device structure and its corresponding properties, we can systematically evaluate the performance of multiple device configurations. Using our framework, we explore three different device configurations with varying oxygen vacancy concentration. For the pristine device, we observed a multi-step filament constriction before full filament rupture. This behavior yields two independent memory states which can be further corroborated by the IV characteristics of the device. For device configurations with a finite oxygen vacancy concentration, a sharp reset profile is observed, which means the IV characteristics of the device would change on modifying the initial structure of HfO$_2$. The inherent stochasticity of the device is further explored through a conductance vs pulse number plot. The device with 0.81\% oxygen vacancies exhibited a linear conductance response with more predictable states for neuromorphic applications. However, for the other two device configurations, the conductance response was non-linear, unpredictable, and stochastic in nature. This stochastic behavior of the device leads us to experience a long standing issue of cycle to cycle and device to device variability.

These findings underscore the central role of local chemistry, interfacial reconstruction, and defect topology in determining emergent functionality in oxide electronic materials. More broadly, our results show that resistive switching in transition-metal-oxide memristors must be understood as a chemically coupled and dynamically evolving process involving both cation and anion transport. This mechanistic insight provides a strong foundation for defect-engineering strategies aimed at improving switching uniformity, multilevel operation, and device reliability, while also highlighting the broader utility of reactive atomistic simulations as predictive tools for the design of functional oxides and electrochemically active materials. Overall, our framework not only provides the physics associated with the device but is also a powerful tool that can simulate different device configurations and perform associated analyses to give a more rational and optimized design of oxide based memristive devices.

\section*{Acknowledgement}This work was supported by DOE Office of Science, Basic Energy Science under the AI-Pathfinder project. Work performed at the Center for Nanoscale Materials, a U.S. Department of Energy Office of Science User Facility, was supported by the U.S. DOE, Office of Basic Energy Sciences, under Contract No. DE-AC02-06CH11357. This work utilized the National Energy Research Scientific Computing Center, a DOE Office of Science User Facility supported by the Office of Science of the US Department of Energy under Contract No. DE-AC02-05CH11231. We also acknowledge the LCRC computing facilities at Argonne.

\printbibliography

%\bibliography{references}
%\bibliographystyle{unsrt}

\newpage
\section*{Supplementary Information}

\subsection*{Detailed Analysis of Functional Form of CTIP}
Electron density of metal atoms are represented by Eq.~\ref{eq:electron_density}

\begin{equation}
\label{eq:electron_density}
\rho_i = \sum_{j=1}^{N} f_i(r_{ij})
\end{equation}

In Eq.~\ref{eq:electron_density} the term f$_i$(r$_{ij}$) represents the electron density of site(represented by Eq.~\ref{eq:site_electron_density}) where atom i will be embedded. This electron density arises from atom j which is at a distance of r$_{ij}$ from atom i.

\begin{equation}
\label{eq:site_electron_density}
f(r) =
\frac{f_e \exp\!\left[-\beta \left(\frac{r}{r_e} - 1\right)\right]}
{1 + \left(\frac{r}{r_e} - \lambda\right)^{20}}
\end{equation}

\begin{equation}
\label{eq:pair_wise_interactions}
\phi(r) =
\frac{A \exp\!\left[-\alpha \left(\frac{r}{r_e} - 1\right)\right]}
{1 + \left(\frac{r}{r_e} - \kappa\right)^{20}}
-
\frac{B \exp\!\left[-\beta \left(\frac{r}{r_e} - 1\right)\right]}
{1 + \left(\frac{r}{r_e} - \lambda\right)^{20}}
\end{equation}

The pairwise term $\phi$$_{ij}$(r$_{ij}$) is the same for both metal-oxygen and oxygen-oxygen interactions but for the electron density it would differ slightly. Hence for oxygen-oxygen interactions the electron density term is represented by Eq.~\ref{eq:oxygen_electron_density}.

\begin{equation}
\label{eq:oxygen_electron_density}
f(r) =
\frac{f_e \exp\!\left[-\gamma \left(\frac{r}{r_e} - 1\right)\right]}
{1 + \left(\frac{r}{r_e} - \nu\right)^{20}}
\end{equation}

The functional form of the embedding term is represented by Eq.~\ref{eq:embedding_functional_form}.

\begin{equation}
\label{eq:embedding_functional_form}
F_j(\rho) =
\sum_{i=0}^{3} F_{ji}
\left( \frac{\rho}{\rho_{e,j}} - 1 \right)^i,
\qquad \rho_{\min,j} \le \rho \le \rho_{\max,j}
\end{equation}

\subsection*{Scaling Factors Used For Analytical Modeling of IV Characteristics from MD Trajectories}

\begin{longtable}{|c|c|c|c|}
\hline
\textbf{Field (V/\AA)} &
\textbf{Voltage (V)} &
\textbf{Scaled-down Voltage (V)} &
\textbf{Adjusted Scaled Voltage (V)} \\ \hline
\endfirsthead

\hline
\textbf{Field (V/\AA)} &
\textbf{Voltage (V)} &
\textbf{Scaled-down Voltage (V)} &
\textbf{Adjusted Scaled Voltage (V)} \\ \hline
\endhead

\hline
\multicolumn{4}{r}{\textit{Continued on next page}} \\ \hline
\endfoot

\endlastfoot

0.0  &   0.00000  &  0.0  &  0.00 \\ \hline
0.2  &  10.75376  &  0.2  &  0.15 \\ \hline
0.4  &  21.50752  &  0.4  &  0.30 \\ \hline
0.6  &  32.26128  &  0.6  &  0.45 \\ \hline
0.8  &  43.01504  &  0.8  &  0.60 \\ \hline
1.0  &  53.76880  &  1.0  &  0.75 \\ \hline
1.2  &  64.52256  &  1.2  &  0.90 \\ \hline
1.4  &  75.27632  &  1.4  &  1.05 \\ \hline
1.2  &  64.52256  &  1.2  &  0.90 \\ \hline
1.0  &  53.76880  &  1.0  &  0.75 \\ \hline
0.8  &  43.01504  &  0.8  &  0.60 \\ \hline
0.6  &  32.26128  &  0.6  &  0.45 \\ \hline
0.4  &  21.50752  &  0.4  &  0.30 \\ \hline
0.2  &  10.75376  &  0.2  &  0.15 \\ \hline
0.0  &   0.00000  &  0.0  &  0.00 \\ \hline
-0.2 & -10.75376  & -0.2 & -0.24 \\ \hline
-0.4 & -21.50752  & -0.4 & -0.48 \\ \hline
-0.6 & -32.26128  & -0.6 & -0.72 \\ \hline
-0.8 & -43.01504  & -0.8 & -0.96 \\ \hline
-1.0 & -53.76880  & -1.0 & -1.20 \\ \hline
-1.2 & -64.52256  & -1.2 & -1.44 \\ \hline
-1.4 & -75.27632  & -1.4 & -1.68 \\ \hline
-1.2 & -64.52256  & -1.2 & -1.44 \\ \hline
-1.0 & -53.76880  & -1.0 & -1.20 \\ \hline
-0.8 & -43.01504  & -0.8 & -0.96 \\ \hline
-0.6 & -32.26128  & -0.6 & -0.72 \\ \hline
-0.4 & -21.50752  & -0.4 & -0.48 \\ \hline
-0.2 & -10.75376  & -0.2 & -0.24 \\ \hline

\caption{The table shows the conversion of electric field to voltage by multiplying the electric field value by 53.7688\AA\ which is the distance between the two electrodes. But for simplicity we scale down the voltage values by 53.7688\AA\ and then apply a scaling factor of 0.75 for set cycle and 1.20 for the reset cycle to adjust the values with experimental data~\cite{Jiang2016Sub10nmTa}.}
\label{tab:field_voltage_updated}
\end{longtable}

\end{document}